# SEISMIC FRAGILITY CURVES FITTING REVISITED: ORDINAL REGRESSION MODELS AND THEIR GENERALIZATION

**Libo Chen**

College of Civil Engineering, Fuzhou University, Fuzhou, Fujian, P.R. China. Email: lbchen@fzu.edu.cn

**Abstract**

This study revisits the modeling of seismic fragility curves by applying ordinal regression models, offering an alternative to the commonly used log-normal distribution function. It compares various ordinal regression approaches, including Cumulative, Sequential, and Adjacent Category models, along with extensions that account for category-specific effects and variance heterogeneity. The methodologies are applied to bridge damage data from the 2008 Wenchuan earthquake, using both frequentist and Bayesian inference methods, with model diagnostics conducted using surrogate residuals. The analysis examines eleven models, from basic forms to those incorporating heteroscedastic extensions and category-specific effects. Based on leave-one-out cross-validation, the Sequential model with category-specific effects performs well compared to traditional Cumulative probit models. The results indicate differences in damage probability predictions between the models, suggesting the potential for more flexible fragility curve modeling techniques to improve seismic risk assessments. This study highlights the importance of continued evaluation of existing methods to enhance the predictive accuracy and applicability of seismic fragility models in performance-based earthquake engineering.

**Keywords**

seismic fragility, ordinal regression models, category-specific effects, variance heterogeneity, Bayesian inference

## 1. INTRODUCTION

In the realm of seismic risk analysis, the formulation of seismic fragility models plays a pivotal role, serving to delineate the probabilistic association between ground motion intensity and structural failure. Seismic fragility is typically represented through either a damage probability matrix or a fragility function. The damage probability matrix[1], an earlier concept, constitutes a tabulation of discrete frequency points denoting structural failure at various ground motion intensities. Conversely, the fragility function[2], widely adopted by researchers over the past four decades, offers a continuous representation of fragility. This function is articulated as the conditional probability that a structure or component will attain or surpass a designated damage state, such as collapse, given a specific ground motion intensity level. Through this lens, the seismic fragility issue can be considered an expansion of the structural reliability challenge, in which the conditional probability of failure is depicted as a function of ground motion intensity.

In seismic fragility research, the lognormal distribution function[3–5] is predominantly selected by most scholars as the foundational model due to its simplicity, widespread acceptance, and the interpretability of its parameters, which enhance the ease of comparing and discussing diverse models. In contrast, alternative distribution functions such as the Logistic[6] and Weibull[7], though present in the literature, may not offer these benefits. The maximum likelihood estimation method is typically employed to determine the parameters of the distribution[8], utilizing sample sets that correlate ground motion intensity with structural failure states. For analytical fragility models, which are thoroughly investigated, alternative statistical methods such as moment estimation and weighted least squares are viable for fitting the fragility function[8–10], contingent upon the analysis approach and data configuration. Additionally, the Bayesian statistical method offers an alternative approach to constructing fragility models. Some scholars have independently developed regression models for seismic capacity or seismic demand, further enabling fragility modeling within the framework of the lognormal distribution function[11,12]. Others have used the Bayesian method to account for statistical uncertainty and the dependence among observations of seismic performance in the fragility modeling of redundant systems[13].

With the advent of Performance-Based Earthquake Engineering (PBEE) over the past three decades[14], seismic fragility research has expanded to include a spectrum of damage states, involving more than just determining the probability of binary outcomes, whether it be failure or non-failure. Seismic fragility models, derived from various data sources, are categorized into empirical, analytical, and expert opinion-based frameworks. Across these categories, the lognormal distribution function consistently underpins the modeling of seismic fragility for different damage states. Employing the maximum likelihood method, Shinozuka (2003) has estimated the parameters of fragility functions for various damage states, either independently or simultaneously, using two distinct modeling approaches based on bridge damage data from earthquakes in California and Japan[15]. The second approach posits that the log standard deviations for the fragility functions should be consistent across all damage states to prevent the crossing of fragility curves, a principle also embraced subsequently by HAZUS[16] and established as the benchmark model for bridge structures. However, this approach raises several questions, including the reasonableness of this fundamental assumption and its potential to restrict the flexibility in representing fragility curves. One might also question the appropriateness of the lognormal distribution function for all fragility curves across various damage states, and whether alternative modeling strategies could be more suitable for the same dataset. Furthermore, the criteria for selecting and evaluating different models warrant careful consideration. This paper seeks to systematically address and discuss these concerns.

In more formal statistical terminology, seismic fragility within the Performance-Based Earthquake Engineering framework can be conceptualized as a probabilistic multinomial classification issue for an ordinal dependent variable, specifically the Damage State (DS), contingent upon certain conditions, such as Intensity Measure (IM) levels. This scenario entails two notable aspects. Firstly, the classification of damage states represents a unique challenge due to their ordinal nature, necessitating the consideration of their sequential order in statistical modeling. While it is possible to disregard this hierarchy and treat damage states as nominal categories using multinomial logistic regression[17], such an approach overlooks the intrinsic order of damage states, potentially leading to information loss and reduced efficacy under specific conditions[18,19]. Secondly, distinct from typical classification tasks that aim to identify the most probable damage category at a given IM level, the primary objective here is to ascertain the likelihood of a structure or component being in each possible damage state. Although fragility models are conventionally expressed through exceedance probabilities, the precise requirement for risk or resilience analysis in practical applications is the probabilities associated with specific intervals delineated by adjacent fragility curves.

To address this issue, a more robust statistical approach involves the use of the ordinal regression method, which is succinctly revisited in this text. Ordinal regression pertains to a spectrum of statistical models that delineate the relationship between an ordered dependent variable and one or more independent variables. The method originated with Aitchison and Silvey (1957), who expanded the binary probit model to accommodate ordinal outcomes encompassing three or more categories[20]. They adopted the concept of a latent continuous variable discernible only at discrete thresholds within the distribution, with the resulting observed variable being ordinal due to the data's incomplete nature. Subsequently, Snell (1964) refined the logit model for ordinal outcomes[21], and Walker and Duncan (1967) implemented the logit model with individual-level data for similar purposes[22]. McKelvey and Zavoina (1975) offered a systematic exposition of the cumulative probit model[23], which extended Aitchison and Silvey's framework to multiple independent variables. McCullagh (1980) introduced the proportional odds model[24], a cumulative approach that models the ordered outcome's cumulative probabilities as a monotonically increasing function transforming a linear predictor to the unit interval. This formulation leads to a probability function congruent with McKelvey and Zavoina's model, rendering it observationally indistinguishable. Both methodologies ignited extensive research on modeling ordered dependent variables—the former primarily within the social sciences, the latter in medical and biostatistical domains.

Prior to 1980, advancements in regression models for ordinal outcomes were confined to cumulative probabilities. Fienberg (1980) presented the Continuation ratio model[25], offering a more adaptable odds ratio structure than the proportional odds model. Tutz (1990) enhanced the Continuation ratio model with a more effective estimation technique[26]. Goodman (1983) proposed the Adjacent Category model[27], renowned for its adaptability and nuanced capture of adjacent category effects. Collectively, these models have demonstrated their utility across diverse fields, including the social sciences, medicine, and engineering. Additionally, the past two decades have seen a steady output of monographs[18,28,29] and reviews[30–32] on ordinal regression, which have enriched the academic discourse. In recent years, scholars have recognized several advantages of the ordinal regression approach over the nominal approach in fragility modeling. These advantages include greater sensitivity to data, parsimony, reduced risk of overfitting, non-crossing fragility curves, and lower associated uncertainty[33]. Furthermore, the ordinal regression approach can be easily extended to fragility models based on multiple ground motion intensity measures, improving the accuracy of the assessment[34]. However, it is important to note that these models remain confined to the construction and discussion of Cumulative models.

This study presents a comprehensive analysis of seismic fragility curve fitting within the ordinal regression framework. By applying various modeling strategies to a single dataset, the research aims to enhance the depth of regression analysis in this field. The paper is structured in three main sections: Section 2 outlines three fundamental ordinal regression models and examines two advanced extensions, evaluating their integration with conventional modeling techniques. This section also addresses key aspects of model fitting and statistical inference, with an emphasis on diagnostic and evaluation procedures. In Section 3, the study uses empirical data from bridge damage during the Wenchuan earthquake for model construction, estimation, diagnostics, evaluation, and selection, employing both frequentist and Bayesian methodologies. This allows for an exploration of the variability in fragility prediction outcomes. Section 4 discusses the historical development of seismic fragility modeling, the challenges of creating analytical fragility models, and provides recommendations for model selection and refinement. The paper concludes by summarizing key findings and insights.

## 2. METHODOLOGY

As highlighted in the introduction, a plethora of ordinal regression models have evolved and found applications across various fields since the 1950s. Notably, their nomenclature has undergone shifts during different historical periods. In broad terms, these models can be categorized into three significant classifications based on their underlying modeling principles. To reference these categories, designations from the existing literature have been employed[32,35], specifically: the Cumulative model, the Sequential model, and the Adjacent Category model.

In this analytical framework, an ordinal response variable is posited, designated as $Y$, which encompasses $K$ ordered categories, ranging from 1 to $K$. The fundamental essence of ordinal regression revolves around the creation of a series of $K-1$ binary classifications, commonly referred to as "cut-point equations" [18]. For each of the three models delineated above, attention is directed toward distinct sets of binary categories, yielding varying probability estimates. Nonetheless, each model integrally accounts for the inherent ordering of the original categorical variables during the modeling endeavor. As such, binary regression models pertaining to the $K-1$ cut-point equations are simultaneously appraised, subject to specific constraints. This exposition delves into the particulars of each of the three models.

*2.1 Basic models*

(1) Cumulative model

The Cumulative model is the most frequently used ordinal regression model, which was typically derived from the assumption that the observed ordinal variable $Y$ originates from the categorization of a latent continuous variable $\tilde{Y}$[28]. The link between the observable categories and the latent variable can be given by $Y=k \Leftrightarrow \tau_{k-1}<\tilde{Y} \leq \tau_k$, where $-\infty=\tau_0<\tau_1<\cdots<\tau_{K-1}<\tau_K=\infty$ are latent thresholds which partition the values of $\tilde{Y}$ into the $K$ ordered categories of $Y$. Let $\tilde{Y}$ follows a linear regression model $\tilde{Y}=\boldsymbol{x}^T\beta+\varepsilon$, where $\varepsilon$ is a noise variable with continuous distribution function $F(\cdot)$, i.e., $F(z)=\mathrm{P}(\varepsilon<z)$. To ensure the identification of the model results in subsequent inferences, the standard deviation of the error term is usually assumed to be 1 and the mean value to be 0.

Combining the above assumptions, the Cumulative model can be obtained as follow:

$$\mathrm{P}(Y \leq k|\boldsymbol{x})=\mathrm{P}\left(\tilde{Y} \leq \tau_k \middle| \boldsymbol{x}\right)=\mathrm{P}(\boldsymbol{x}^T\beta+\varepsilon \leq \tau_k)=\mathrm{P}(\varepsilon \leq \tau_k-\boldsymbol{x}^T\beta)=F(\tau_k-\boldsymbol{x}^T\beta), \qquad k=1,2\ldots,K-1 \quad (1)$$

Further, the probabilities of occurrence (i.e., of being in a particular category) can be easily derived as follow:

$$\mathrm{P}(Y=k|\boldsymbol{x})=\mathrm{P}(Y \leq k|\boldsymbol{x})-\mathrm{P}(Y \leq k-1|\boldsymbol{x})=F(\tau_k-\boldsymbol{x}^T\beta)-F(\tau_{k-1}-\boldsymbol{x}^T\beta), \qquad k=1,2\ldots,K \quad (2)$$

The specific form of $F$ relies on the assumed distribution of the error term, and the common options include logistic, normal and extreme value distribution, which associated with the logit, probit and cloglog link functions (the inverse distribution functions $F^{-1}$) typically. For consistency with traditional conventions in seismic fragility analysis, the probit function is selected as the link function in this paper.

In the context of the seismic fragility problem, the ordinal variable corresponds to the damage state (DS), where conventionally, no damage, minor damage, moderate damage, major damage, and complete damage are denoted by $k=1,2,3,4,5$, respectively. Additionally, the logarithm of a scaled ground motion intensity measure, $\ln(IM)$, is chosen as the independent variable. With the derivation outlined above, the probability of attaining a specific damage state equal to or less than a given ground motion intensity $IM=x$ can be obtained.

$$\mathrm{P}(DS \leq k|IM=x)=\mathrm{P}(\beta\ln(x)+\varepsilon \leq \tau_k)=\Phi(\tau_k-\beta\ln(x)) \quad (3)$$

Notice that the traditional seismic fragility is usually defined as the probability of exceeding a given damage state. Due to the symmetry of the normal distribution, it is easy to obtain the fragility curves in the form of lognormal cumulative distribution function with two parameters $\theta_k$ and $\tilde{\beta}$ by simple parameter conversions

$$\mathrm{Fr}_k(IM=x)=\mathrm{P}(DS>k|IM=x)=1-\mathrm{P}(DS \leq k|IM=x)=\Phi(\beta\ln(x)-\tau_k)=\Phi\left(\frac{\ln\left(\frac{x}{\theta_k}\right)}{\tilde{\beta}}\right) \quad (4)$$

where $\theta_k=\exp(\tau_k/\beta)$ is the median value and $\tilde{\beta}=1/\beta$ is the logarithmic standard deviation of the normal distribution. The physical meaning of the median value is the value of the ground motion intensity measure corresponding to a failure exceedance probability of 50%, while the logarithmic standard deviation portrays the overall flatness of the fragility curve. In cases where two fragility functions possess identical logarithmic standard deviations, a higher median value corresponds to a reduced exceedance probability of structural damage under the same ground motion intensity, which indicates that it is less fragile. Likewise, when comparing two fragility curves with identical median values that intersect at this median, the curve with a larger logarithmic standard deviation will be flatter. This characteristic implies that it has a higher exceedance probability of damage at lower ground motion intensities and a lower exceedance probability at higher intensities.

As stated in the introduction, the Equation (4) represents the conventional expression in current seismic fragility studies. However, its derivation reveals that utilizing the lognormal distribution function simply presents another form of the Cumulative probit model, given the assumptions mentioned earlier.

(2) Sequential model

The Sequential model can be derived from the assumption that $1,\ldots,k$ are reached successively. It might reflect the successive transition to higher categories in a stepwise model, i.e., in order to achieve a category $k$, one has to first achieve all lower categories 1 to $k-1$ [32]. For every category $k\in\{1,\ldots,K\}$ there is a latent continuous variable $\tilde{Y}_k$ determining the transition between the $k$th and the $k+1$th category.

The Sequential model can be executed as follows: Let the process start in category 1. First, it is checked whether $\tilde{Y}_1$ surpasses the first threshold $\tau_1$: if $\tilde{Y}_1\leq\tau_1$, the process stops and the result is $Y=1$; else if $\tilde{Y}_1>\tau_1$, at least category 2 is achieved (i.e., $Y>1$) and the process continuous. Second, it is checked whether $\tilde{Y}_2$ surpasses threshold $\tau_2$: if $\tilde{Y}_2\leq\tau_2$, the process stops with the result $Y=2$; else, $\tilde{Y}_2>\tau_2$, the process continuous with at least category 3 (i.e., $Y>2$), and so on in a similar fashion.

In the same way as for the Cumulative model, assume that $\tilde{Y}_k$ depends on the predictor $\boldsymbol{x}$ and error $\varepsilon_k$: $\tilde{Y}_k=\boldsymbol{x}^T\beta+\varepsilon_k$, where $\varepsilon_k$ has mean zero and is distributed according to $F(\cdot)$: $F(z)=\mathrm{P}(\varepsilon_k<z)$. The decision between category $\{k\}$ and categories $\{k+1,\ldots,K\}$ in the $k$th step can be modeled by the binary model

$$\mathrm{P}(Y=k|Y\geq k,\boldsymbol{x})=\mathrm{P}\left(\tilde{Y}_k\leq\tau_k\middle|\boldsymbol{x}\right)=\mathrm{P}(\boldsymbol{x}^T\beta+\varepsilon_k\leq\tau_k)=\mathrm{P}(\varepsilon_k\leq\tau_k-\boldsymbol{x}^T\beta)=F(\tau_k-\boldsymbol{x}^T\beta),k=1,2\ldots,K-1 \quad (5)$$

Further, the unconditional probabilities of occurrence in the Sequential model are given by

$$\mathrm{P}(Y=k|\boldsymbol{x})=\mathrm{P}(Y=k|Y\geq k,\boldsymbol{x})\prod_{j=1}^{k-1}\mathrm{P}(Y>j|Y\geq j,\boldsymbol{x})=F(\tau_k-\boldsymbol{x}^T\beta)\prod_{j=1}^{k-1}\left(1-F\left(\tau_j-\boldsymbol{x}^T\beta\right)\right),k=1,2\ldots,K \quad (6)$$

It needs to be clarified that not all ordinal variables can be modeled using the Sequential approach, such as a Likert-type scale in social surveys (e.g. strongly disagree, disagree, undecided, agree, and strongly agree), since there are no preceding step processes between each other as previously described, i.e. it is not necessary to agree with a statement before disagreeing with it. But for some other variables, this choice is natural, such as educational attainment (e.g. less than high school, high school or equivalent, associate or junior college, college, and postgraduate).

Fortunately, for the damage state in seismic fragility problem, it is a reasonable and intuitive approach. This is because the damage states of a structure under an earthquake usually can be considered to be achieved in a through-sequence, and it is natural to believe that structures in a more severely damaged state constitute a subset of those in a state of less severe damage. Consequently, the probabilities of more severe states can be determined by taking into consideration that they are statistically conditional to the probabilities associated with less severe states of damage.

Based on the previous assumptions and the above derivation, we can obtain the probability of being equal to a specific damage state at a given ground motion intensity $IM=x$

$$\mathrm{P}(DS=k|IM=x)=\Phi(\tau_k-\beta\ln(x))\prod_{j=1}^{k-1}\left(1-\Phi\left(\tau_j-\beta\ln(x)\right)\right) \quad (7)$$

The fragility curves for different damage states are given by

$$\mathrm{Fr}_k(IM = x) = \mathrm{P}(DS > k|IM = x) = 1 - \sum_{j=1}^{k} \mathrm{P}(DS = j|IM = x) \quad (8)$$

Furthermore, alongside the aforementioned derivation process, an alternative approach for constructing the Sequential model is elaborated upon in the first point of the discussion section (historical notes).

(3) Adjacent Category model

The Adjacent Category model is an ordinal regression approach that focuses on adjacent probability, which is the likelihood of a particular category $k$ in relation to its immediate successor, category $k+1$ [27]. This model is distinctive in its emphasis on pairwise comparisons between consecutive outcome categories within the binary cut-point equations, a feature not present in other ordinal regression models. This makes the Adjacent Category model especially suitable for analyzing ordinal outcomes where the individual categories hold substantive interest for researchers. Its strength lies in the ability to directly compare discrete categories, rather than relying on distinct points along a cumulative distribution[18]. Additionally, the Adjacent Category model is well-suited for datasets with unbalanced distributions or small sample sizes[36], offering flexibility across different research contexts. By focusing on adjacent comparisons, this model provides deeper insights into the transitions between ordinal levels, offering a more detailed understanding of the underlying ordinal structure.

Within the framework of the Adjacent Category model, the distinction between category $k$ rather than category $k+1$ is achieved using latent variables $\tilde{Y}_k$ with the thresholds $\tau_k$ and cumulative distribution function $F(\cdot)$, the basic form can be written as follow

$$\mathrm{P}(Y = k|Y \in \{k, k+1\}, \boldsymbol{x}) = F(\tau_k - \boldsymbol{x}^T\beta), \qquad k = 1,2 \ldots, K-1 \quad (9)$$

The unconditional probabilities of occurrence in the Adjacent Category model can be given by [35]

$$\mathrm{P}(Y = k|\boldsymbol{x}) = \frac{\prod_{j=1}^{k-1}\left(1 - F\left(\tau_j - \boldsymbol{x}^T\beta\right)\right)\prod_{j=k}^{K} F\left(\tau_j - \boldsymbol{x}^T\beta\right)}{\sum_{r=1}^{K+1}\left(\prod_{j=1}^{r-1}\left(1 - F\left(\tau_j - \boldsymbol{x}^T\beta\right)\right)\prod_{j=r}^{K} F\left(\tau_j - \boldsymbol{x}^T\beta\right)\right)}, \qquad k = 1,2 \ldots, K \quad (10)$$

with $\prod_{j=1}^{0}\left(1 - F\left(\tau_j - \boldsymbol{x}^T\beta\right)\right) = \prod_{j=K+1}^{K} F\left(\tau_j - \boldsymbol{x}^T\beta\right) := 1$ for notational convenience. For a more detailed derivation of this formula, see the Appendix.

In previous academic literature, the majority of scholars opted to employ the logistic distribution function when applying the Adjacent Category model, often motivated by a desire for simplification. In this context, logits are built locally for adjacent categories of the form

$$\log\left(\frac{\mathrm{P}(Y = k|\boldsymbol{x})}{\mathrm{P}(Y = k+1|\boldsymbol{x})}\right) = \tau_k - \boldsymbol{x}^T\beta, \qquad k = 1,2 \ldots, K-1 \quad (11)$$

where the predicted probability for a given category $k$ is given by the following equations

$$\mathrm{P}(Y = k|\boldsymbol{x}) = \frac{\exp\left(\sum_{j=k}^{K}\left(\tau_j - \boldsymbol{x}^T\beta\right)\right)}{\sum_{r=1}^{K+1}\left(\exp\left(\sum_{j=r}^{K}\left(\tau_j - \boldsymbol{x}^T\beta\right)\right)\right)}, \qquad k = 1,2 \ldots, K \quad (12)$$

with $\exp\left(\sum_{j=K+1}^{K}\left(\tau_j - \boldsymbol{x}^T\beta\right)\right) := 1$ for convenience.

As previously noted, for the sake of contextual coherence and to minimize the interference of different choices of link functions on the results, the probit function is employed to construct the seismic fragility model in this study. The probabilities of being equal to a specific damage state at a given ground motion intensity $IM = x$ are given by

$$\mathrm{P}(DS=k|IM=x)=\frac{\prod_{j=1}^{k-1}\left(1-\Phi\left(\tau_j-\beta\ln(x)\right)\right)\prod_{j=k}^{K}\Phi\left(\tau_j-\beta\ln(x)\right)}{\sum_{r=1}^{K+1}\left(\prod_{j=1}^{r-1}\left(1-\Phi\left(\tau_j-\beta\ln(x)\right)\right)\prod_{j=r}^{K}\Phi\left(\tau_j-\beta\ln(x)\right)\right)} \tag{13}$$

The formulation of the fragility curves for different damage states is identical to Equation (8).

In summary, the analysis of the three discussed models reveals that they fundamentally rely on a binary model framework. The primary difference lies in how these binary elements are integrated into a comprehensive model that effectively captures the ordinal relationships among categories. This differentiation not only facilitates a clear classification but also sheds light on the models' operational dynamics: within the Cumulative model, binary models act as unconditional models across various category groups, i.e., $\{1,2,\dots,k\}$ **vs** $\{k+1,k+2,\dots,K\}$. In contrast, Sequential model perform conditional comparisons of both individual categories and category groups, i.e., $\{k\}$ **vs** $\{k+1,k+2,\dots,K\}$, while Adjacent Category model conduct conditional comparisons of individual categories, i.e., $\{k\}$ **vs** $\{k+1\}$.

It is essential to highlight that in constructing fragility models, the utilization of Cumulative models is notably straightforward and uncomplicated. Similar to the lognormal distribution function, the Logistic and Weibull functional forms, as cited in the literature and mentioned earlier, represent distinct variants of Cumulative models based on different assumptions regarding the link functions. In contrast, Sequential and Adjacent Category models derive the probabilities for each damage state directly, with exceedance (or cumulative) probabilities calculated through accumulation. Consequently, the seismic fragility functions derived from these two models deviate from the simplistic structure of standard distribution functions. However, despite the added complexity of the Sequential and Adjacent Category models compared to the Cumulative model, they offer enhanced flexibility and performance in specific contexts, details of which will be elaborated further.

### *2.2 Generalizations of ordinal models*

(1) Category-specific effects

In all of the above ordinal models, it's by default assumed that the effect of predictors captured in $\boldsymbol{x}^T\beta$ is the same for the category (or group of categories) being compared, which makes the models simple and allows easy interpretation of the regression parameters. This assumption is often referred to as the parallel regression assumption[18] (also known as the proportional odds assumption for Cumulative logit model specifically[24]), however, it may not be a reasonable assumption in some cases. It is often possible that a predictor has different impacts for different response categories of $Y$. For the three models mentioned above, the interpretation of their specific meanings differs. Take the Adjacent Category model as an example, the following specific scenarios may occur in practice: the intensity measure of ground motion may have little relation to whether the structure in minor damage over in moderate damage, but strongly predict whether major damage is preferred over complete damage.

There are a variety of different tests for the parallel regression assumption, such as likelihood ratio test, the Wald test, or the score test[18]. In addition to formal tests of statistical significance, supplementary informal assessments based on residual analysis or model comparisons utilizing information criteria can also prove valuable in the analysis. For scenarios where the assumption is not satisfied, a feasible generalization of the model is to take the category-specific effects into account by estimating $K-1$ slope coefficients rather than one coefficient for the predictors, i.e., replace $\boldsymbol{x}^T\beta$ by $\boldsymbol{x}^T\beta_k$. However, for the Cumulative model, this operation is equivalent to transforming the original model into a series of binary regression models with separate parameters for different categories' groupings $\{1,\dots,k\},\{k+1,\dots,K\}$. It also implies the logical failure of a single latent variable. Although the flexibility of the model was improved, the results is not always reasonable, because it may result in negative probabilities.

Specifically for the seismic fragility problem, the estimation results of the separate parameters may lead to the crossover of the fragility curves, similar to the situation of the first method by Shinozuka[15], which do not follow the common sense. And this is why a reluctant compromise that had to be made in the second method, i.e., the identical logarithmic standard deviation of the fragility curves corresponding to different damage states is assumed, which also reverts to the basic cumulative model, as the Equation (4) presented.

It is worth mentioning that for the case of intersection of fragility curves, although some targeted correction strategies have been proposed[37], however, these are remedial measures in retrospect, which are not good modeling strategies from the authors' point of view. Had we been able to predict its occurrence in advance, alternative modeling strategies could have been employed to circumvent it. For the Sequential model or Adjacent Category model, incorporating category-specific effects will not cause any problems. The underlying reason for this is that both models are fundamentally conditional probability models, and this factor may be taken into consideration during the conversion from conditional to unconditional probability calculations.

(2) Variance heterogeneity

The classical linear regression assumes that error terms are homoscedastic. If the assumption is violated, i.e., error variances are heteroskedastic, then the estimates of variable coefficients remain unbiased, but the standard error terms and significance tests will be distorted. In the real world, there are many causes of the heteroscedasticity phenomenon. One of the possible scenarios is variable omissions, that is, the residual includes some variables that are linearly related to some independent variables, but are not considered by the model, resulting in the correlation between the residuals and independent variables. In fact, the heteroscedasticity phenomenon is often noted in existing probabilistic seismic demand analyses [38,39], and the most likely explanation is that a scalar ground motion intensity measure cannot fully characterize all the features of the ground motion input, and some of the missing features may contribute to the seismic demand of the structural component. A natural association is whether similar scenarios would extrapolated to influence the construction of seismic fragility models?

For the above-mentioned ordinal regression models, it is also by default assumed that the variance of the latent variable to be the same throughout the model. However, this may not be the case and the consequences of heteroscedasticity can be much greater: coefficient estimates can be biased and cross-group comparisons in particular can be misleading[40]. If variance heterogeneity exists, one way to introduce it is to further model the unexplained heterogeneities with re-parameterization, i.e. model it explicitly as depending on covariates, which is known as the location-scale model[24], and is also referred to as the heterogeneous choice model in sociological monographs[41].

Taking the Cumulative model as an example, if the underlying latent variable is given by $\tilde{Y} = \boldsymbol{x}^T\beta + \varepsilon\sigma$ with $\sigma = \exp(\boldsymbol{z}^T\gamma)$ where $\boldsymbol{z}$ is an additional vector of covariates, we can obtains the location-scale model as follow:

$$\mathrm{P}(Y \le k|\boldsymbol{x}) = F\left(\frac{\tau_k - \boldsymbol{x}^T\beta}{\exp(\boldsymbol{z}^T\gamma)}\right), \qquad k = 1, \dots, K-1 \tag{14}$$

In the given model, the $\boldsymbol{x}$ variables function as explanatory factors and are considered to be the determinants of the outcome or choice; whereas the $\boldsymbol{z}$ variables delineate groups possessing distinct error variances within the latent variable framework. It is not mandatory for the $\boldsymbol{z}$ and $\boldsymbol{x}$ variables to overlap, although such an intersection is permissible. The vectors $\beta$ and $\gamma$ represent coefficients indicating the influence of the $\boldsymbol{x}$ variables on the choice and the impact of the $\boldsymbol{z}$ variables on the variance, respectively-or more precisely, on the logarithm of $\sigma$. In this study, both $\boldsymbol{x}$ and $\boldsymbol{z}$ variables are defined as the logarithm of a scale ground motion intensity measure $\ln(IM)$.

In the aforementioned expression, the numerator is denominated as the location equation, while the denominator is designated as the scale equation. These can also be termed the choice and variance equations, correspondingly. It is notable that the location equation encompasses a constant term, whereas the scale equation does not. Traditional probit model, which lack a variance equation, are particular instances of the general model described here, characterized by a constant $\sigma = 1$ across all observations.

*2.3 Inference for ordinal regression models*

Traditionally, an ordinal regression model can be estimated using the maximum likelihood method [28], which involves a series of steps that integrate the conceptual framework of ordinal regression with iterative numerical methods for optimization. Consider a sample of $n$ individuals. To formulate the likelihood function, the original response variable $y$ is represented as a vector of binary responses. For each individual $i$, let $\boldsymbol{y}_i = (y_{i1}, \dots, y_{iK})$ be a vector of binary responses $y_{ij}$, with $y_{ij} = 1$ if the response is in category $j$ and $y_{ij} = 0$ otherwise $(1 \leq j \leq K)$. The values of the explanatory variables for subject $i$ are denoted by $\boldsymbol{x}_i$, and $\pi_j(\boldsymbol{x}_i)$ represents $P(Y_i = j \mid \boldsymbol{X} = \boldsymbol{x}_i)$. The specific expressions for $\pi_j(\boldsymbol{x}_i)$ in the three fundamental models outlined in the preceding section are presented in Equations (2), (6), and (10). For independent observations, the likelihood function derives from the product of the multinomial mass functions for the $n$ subjects is

$$L(\boldsymbol{\theta}) = \prod_{i=1}^{n} \left[ \prod_{j=1}^{K} \left[\pi_j(\boldsymbol{x}_i)\right]^{y_{ij}} \right] \tag{15}$$

The log-likelihood, which is often easier to work with, is given by

$$l(\boldsymbol{\theta}) = \sum_{i=1}^{n} \left[ \sum_{j=1}^{K} y_{ij} \log \pi_j(\boldsymbol{x}_i) \right] \tag{16}$$

Since the equations are nonlinear in parameters and do not have a closed-form solution, iterative methods like the Fisher scoring or Newton-Raphson method are used to numerically obtain the maximum likelihood estimate [22,24]. Several R packages are available for fitting ordinal regression models, including the highly recommended ‘MASS’, as well as ‘VGAM’, ‘ordinal’, and ‘rms’. While each package has its unique focus, they predominantly serve the Cumulative model. Additionally, some of these packages offer modeling options for Sequential and Adjacent Category models, as well as the previously mentioned two extensions.

In addition to the traditional frequentist statistical inference method, the Bayesian approach has gained increasing popularity in recent years[42–44]. This alternative method employs probability distributions for both parameters and data, assuming a prior distribution for parameters that may incorporate prior beliefs or existing information about their values. This prior knowledge is integrated with the data-derived information through the likelihood function, culminating in a posterior distribution for the parameters. Distinct from the frequentist approach, Bayesian inference excels in managing small sample datasets, integrating complex hierarchical models, and leveraging prior knowledge. Additionally, Bayesian inference provides a measure of uncertainty via the posterior distribution, offering a more comprehensive uncertainty estimation than frequentist estimates. In terms of practical implementation, Bayesian methods typically employ Markov Chain Monte Carlo (MCMC) techniques, including Metropolis-Hastings updates, Gibbs sampling, and Hamiltonian Monte Carlo, to fit models. The implementation of these methods is well-supported across the platform R, which providing extensive Bayesian analysis packages and libraries, such as JAGS, NIMBLE, and Stan.

*2.4 Goodness-of-fit and model evaluation*

(1) Residuals and diagnostics for ordinal regressions

A persistent challenge in ordinal regression analysis is the absence of effective tools for validating model assumptions. This issue stems from the nature of ordinal variables, which have discrete, non-numerical labels representing ordered categories. Liu and Zhang (2018) introduced a novel residual type [45], derived from a continuous surrogate variable $S$ for the ordinal outcome $Y$. This residual demonstrates null properties akin to those found in common residuals for continuous outcomes and is useful in identifying model misspecifications, including incorrect link functions, heteroscedasticity, and issues with proportionality. For a comprehensive understanding and practical application, the fundamental principles of this method are discussed here.

This surrogate residual is defined as

$$R_s = S - E(S|\boldsymbol{X}) \tag{17}$$

where $S$ is a continuous variable derived from the conditional distribution of the latent variable $Z$ given $Y$. Specifically, the authors demonstrated that, for a given $Y = y$, $S$ adheres to a truncated distribution, resulting from the truncation of $Z$'s distribution within a defined interval. The advantage of using the surrogate residual $R_s$ lies in its continuity, as it is based on the continuous variable $S$. Moreover, if the hypothesized model corresponds accurately with the actual model, then $R_s$ will exhibit certain characteristics:

(a) (Symmetry around zero) $E(R_s|\boldsymbol{X}) = 0$

(b) (Homogeneous variance) $Var(R_s|\boldsymbol{X})$ is a constant, not depending on $\boldsymbol{X}$

(c) (Explicit reference distribution) the empirical distribution of $R_s$ approximates an explicit distribution that is related to the link function $F^{-1}(\cdot)$. In particular, independent of $\boldsymbol{X}$, $R_s \sim F(c + \int u dF(u))$, where $c$ is a constant. According to property (a), if $\int u dF(u) = 0$, then $R_s \sim F(\cdot)$.

Properties (a-c) allow for a thorough examination of the residuals to check model adequacy and misspecification of the link function. In the course of specific practices, we can define a surrogate using a technique called jittering, and then got the surrogate residuals in the same way as Equation (17). Further we can check if the hypothesized model is correct by plotting the relevant residuals' diagrams.

Building upon this groundwork, the 'sure' software package has been developed in the R programming environment [46]. It is compatible with a wide array of R packages for fitting Cumulative link models and other model types, significantly facilitating the diagnosis and evaluation of ordinal regression outcomes.

(2) Model evaluation and selection

In the traditional approach to ordinal regression, model evaluation and selection heavily rely on pseudo-$R^2$ statistics [19] and information-theoretic measures such as the Akaike Information Criterion (AIC) and the Bayesian Information Criterion (BIC). Pseudo-$R^2$s, although not directly comparable to R-squared in linear regression, provide a measure of model fit relative to a null model, with different versions offering varied perspectives on the data's explained variance. The common Pseudo-R2s include McFadden's $R^2$, Cox & Snell's $R^2$, McKelvey & Zavoina's $R^2$, etc. The AIC and BIC balance the model's goodness of fit with its complexity, where lower values indicate a more preferable model. AIC focuses on the trade-off between the fit and the number of parameters, while BIC adds a stricter penalty for models with more parameters, aligning more closely with Bayesian probability principles.

The Bayesian approach, on the other hand, incorporates more comprehensive methods like the Deviance Information Criterion (DIC) [47] and the Widely Applicable Information Criterion (WAIC) [48] for model evaluation in ordinal regression. DIC, akin to a Bayesian version of AIC, assesses model quality by balancing fit and complexity, including a penalty based on the effective number of parameters. Specifically, DIC is calculated as the sum of the posterior mean of the deviance and a penalty term for model complexity. WAIC, another advanced tool, evaluates models based on the log pointwise predictive density (lppd), corrected for overfitting. It is computed by summing the logarithm of the average predictive probability for each observation and subtracting a penalty that increases with model complexity, measured by the variance of the log predictive density across the posterior. This approach offers a more nuanced assessment, especially for complex models or those with latent variables.

Recent scholarship has introduced an alternative evaluation metric based on leave-one-out cross-validation (LOOCV) [49]. This method gauges a model's predictive accuracy by iteratively excluding each data point to evaluate prediction quality for the omitted point. Unlike WAIC, LOO does not explicitly incorporate a parameter penalty term, yet it indirectly addresses model complexity during the cross-validation process. For model comparison, LOO assesses predictive accuracy by excluding and predicting one observation at a time. Models yielding the highest expected log pointwise predictive density (elpd_loo) are favored, as this approach naturally penalizes complexity and overfitting. Notably, PSIS-LOO, a variant employing Pareto-smoothed importance sampling, demonstrates enhanced robustness in finite-sample scenarios, particularly with weak priors or influential observations. This robustness stems from importance weights regularization in PSIS. The 'loo' R package implementation of these methods simplifies their application in statistical practice and provides approximate standard errors for predictive error estimates and model comparisons.

## 3. CASE STUDY

### *3.1 Damage data profiles*

The seismic dataset utilized in this paper is sourced from the empirical bridge damage data from the 2008 Wenchuan earthquake [50]. On May 12, 2008, at 14:28 local time, a devastating earthquake with a magnitude of Ms 8.0 struck in close proximity to Yingxiu, Wenchuan County, Sichuan Province, China. This catastrophic event inflicted substantial damage upon a multitude of highway bridges, primarily concentrated in the western regions of Sichuan Province and the southern sectors of Gansu and Shaanxi Provinces. The extent of earthquake-induced damage to highway bridges correlates closely with the seismic dynamics of the fault responsible for the earthquake and the fault's strike direction, demonstrating a consistent spatial distribution pattern across various levels of seismic intensity. The most severely impacted highway bridges are concentrated within three routes near the epicenter in Yingxiu and three additional routes near Beichuan.

In this study, a systematic and comprehensive categorization and statistical analysis is undertake for the bridges along these six aforementioned routes, in addition to the Duwen Expressway, which was under construction at the time. This extensive dataset encompasses a total of 442 bridge samples. The bridge damage is classified into five distinct levels according to the extent of loss of load-bearing capacity, denoted as A0, A, B, C, and D, indicates the none damage, minor damage, moderate damage, major damage, and complete damage respectively, based on the 'Investigation Report on Seismic Damage in Highway Engineering from the Wenchuan Earthquake' [51].

Within the context of this research, the Peak Ground Acceleration (PGA) is selected as the primary indicator of ground motion intensity for fragility analysis of the bridges affected by the Wenchuan earthquake. Prior scholarly works have frequently employed interpolated PGA values derived from Shake Maps at the bridge locations for further analysis. However, this study initiates by compiling and summarizing data from 131 seismic stations located in Sichuan Province. The investigation reveals that utilizing interpolation method by the shake map to estimate PGA values at bridge sites can yield significant errors. This discrepancy is predominantly attributed to the limited number of seismic stations and their sparse distribution across the region. Consequently, this study employs a ground motion attenuation model to estimate PGA as a measure of ground motion intensity.

The attenuation models are founded on empirical relationships derived from statistical regressions of historical seismic data. For instance, Zhao et al. (2006) developed a spectral acceleration attenuation model based on extensive earthquake records in Japan [52]. Subsequently, in 2010, Lu et al. conducted a comprehensive analysis of ground motion prediction models for the Wenchuan earthquake [53], ultimately confirming Zhao's model as the most accurate in predicting peak acceleration and short-period spectral acceleration in the near-field areas of the Wenchuan earthquake when compared to the Next Generation Attenuation (NGA) models proposed by other scholars. In this study, Zhao's model is selected as the foundational model for predicting the ground motion intensity measure PGA. Subsequently, it is refined through modifications based on measured data from the Wenchuan earthquake to obtain precise PGA values at specific locations of interest.

Upon estimating the PGA values at each bridge site, we combined this data with the damage states of bridges from prior investigation to create the IM - DS sample set for damaged bridges affected by the Wenchuan earthquake, as illustrated in Figure 1. It is important to acknowledge that in empirical seismic fragility modeling, the inherent uncertainty in the estimated ground motion intensities (rather than directly measured) can impact the fragility model's outcomes [9]. Relevant literature exists on this topic [54], also indicating that such uncertainties are more effectively addressed within a Bayesian framework. However, this paper's focus is on the evaluation and selection of the seismic fragility model itself, based on a consistent dataset. To avoid excessive length and digression, the influence of this uncertainty is not addressed in the analysis.

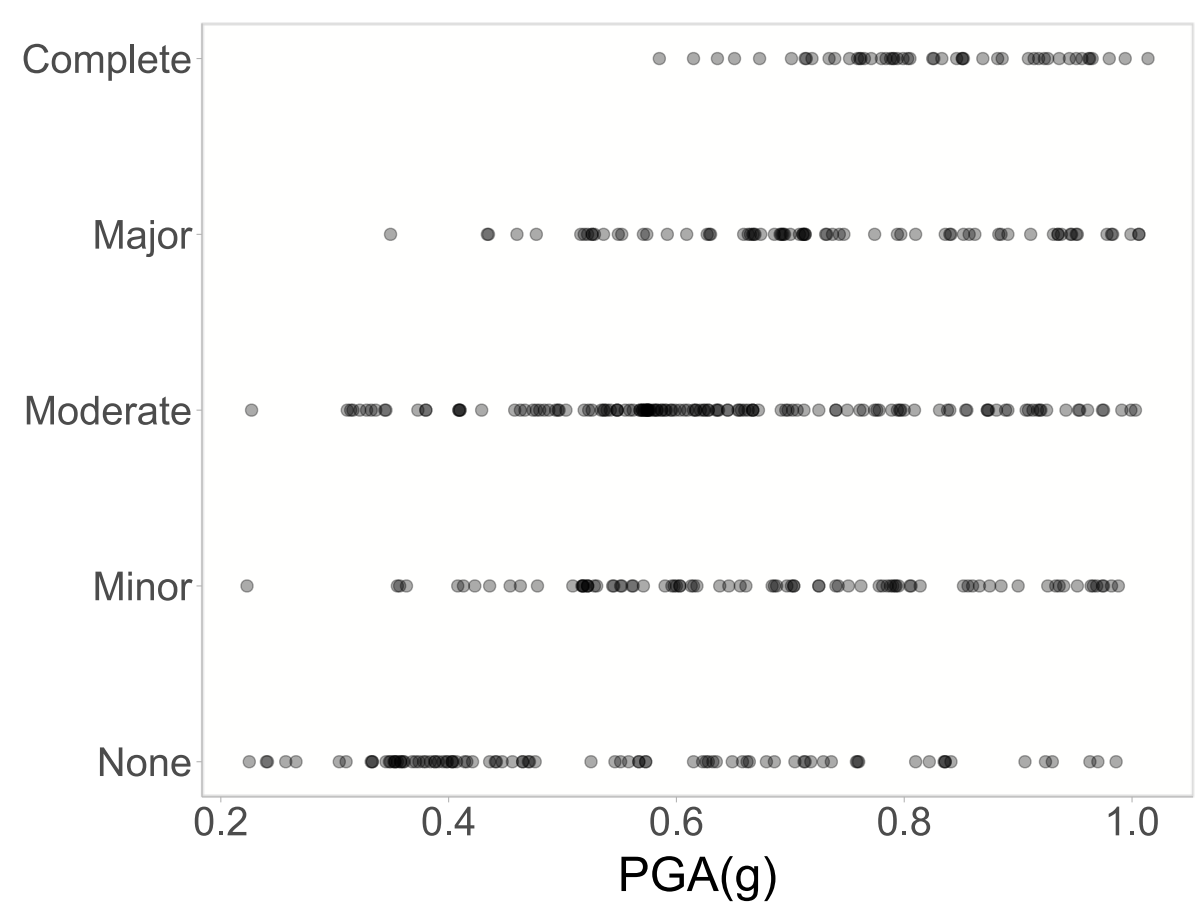

Figure 1 Scatterplot of IM - DS sample set for damaged bridges affected by Wenchuan earthquake (The study comprised 442 samples, distributed across five damage states: 91 with no damage, 84 with minor damage, 142 with moderate damage, 74 with major damage, and 51 with complete damage)

### *3.2 Seismic fragility modelling*

(1) Residuals and diagnostics for cumulative probit regression

Prior to delving into a comparative analysis of various modeling strategies, this study initially applies the traditional Cumulative probit regression method to the dataset, supplemented by the surrogate residual method outlined in Section 2.4 for diagnostic purposes. The 'polr' function from the MASS package is employed for regression [55], and the results are integrated with the 'sure' package to produce surrogate residuals for graphical representation. This approach facilitates an examination of the validity of assumptions such as heteroscedasticity and proportionality.

The analysis begins with an examination of scatter plots between surrogate residuals and independent variables,

as depicted in Figure 2(a). The residuals predominantly cluster above and below the x-axis, and the nonparametric smooth trend line (depicted in red) remains close to the zero value without displaying any discernible trend. This suggests that the mean structure does not exhibit notable nonlinearity, thereby supporting the validity of the linear function assumption. Additionally, an increase in the variance of surrogate residuals with larger independent variable values hints at mild heteroscedasticity in the data. Furthermore, the QQ plot of the residuals, also shown in Figure 2(b), aligns closely with the diagonal line without significant deviation. This alignment corroborates the appropriateness of using a probit link function for this dataset.

The parallel regression assumption, elaborated upon in Section 2.2, posits consistent mean structures (specifically, $\beta\ln\mathrm{PGA}$ in this study) across each category. To evaluate this assumption, the study bifurcated the dataset into two subsets and conducted separate analyses using a Cumulative probit regression model. The first subset included samples representing the initial three damage states: None damage, Minor damage, and Moderate damage. Conversely, the second subset encompassed samples from the latter three damage states: Moderate damage, Major damage, and Complete damage. The respective models for these subsets are described as follows:

$$\begin{cases} \mathrm{P}(DS \le k|PGA) = \Phi(\tau_k - \beta_1\ln\mathrm{PGA}), k = 1,2,3 \\ \mathrm{P}(DS \le k|PGA) = \Phi(\tau_k - \beta_2\ln\mathrm{PGA}), k = 3,4,5 \end{cases} \tag{18}$$

The assessment of the parallel regression assumption involves verifying the equality of coefficients $\beta_1$ and $\beta_2$ derived from confirmatory regression, essentially testing the validity of the equation $\beta_1 - \beta_2 = 0$. Drawing on prior research, this study generates surrogates for damage states $S_1 \sim N(-\beta_1\ln\mathrm{PGA}, 1)$ and $S_2 \sim N(-\beta_2\ln\mathrm{PGA}, 1)$, both conditional on $\ln\mathrm{PGA}$. Then the difference $D = S_2 - S_1$ can be defined, which also conditional on $\ln\mathrm{PGA}$ and follows a normal distribution $N\big((\beta_1 - \beta_2)\ln\mathrm{PGA}, 2\big)$. If the equation $\beta_1 - \beta_2 = 0$ holds, $D$ should be independent of $\ln\mathrm{PGA}$, which can be easily checked by plotting $D$ against $\ln\mathrm{PGA}$. A scatter plot analysis, illustrated in Figure 2(c), reveals a discernible trend with the nonparametric smooth trend line (depicted in red) markedly deviating from zero. This deviation indicates the equation's invalidity, thereby challenging the parallel lines hypothesis.

In summary, the residuals analysis of the Cumulative probit model using the Wenchuan earthquake bridge damage dataset suggests that the assumptions regarding mean structure and the probit link function—i.e., linearity concerning independent variables and normality of latent variables—generally align with expectations. However, it is important to note the presence of slight heteroscedasticity and the failure to meet the parallel regression assumption. These findings underscore the potential value of exploring alternative modeling strategies in subsequent research.

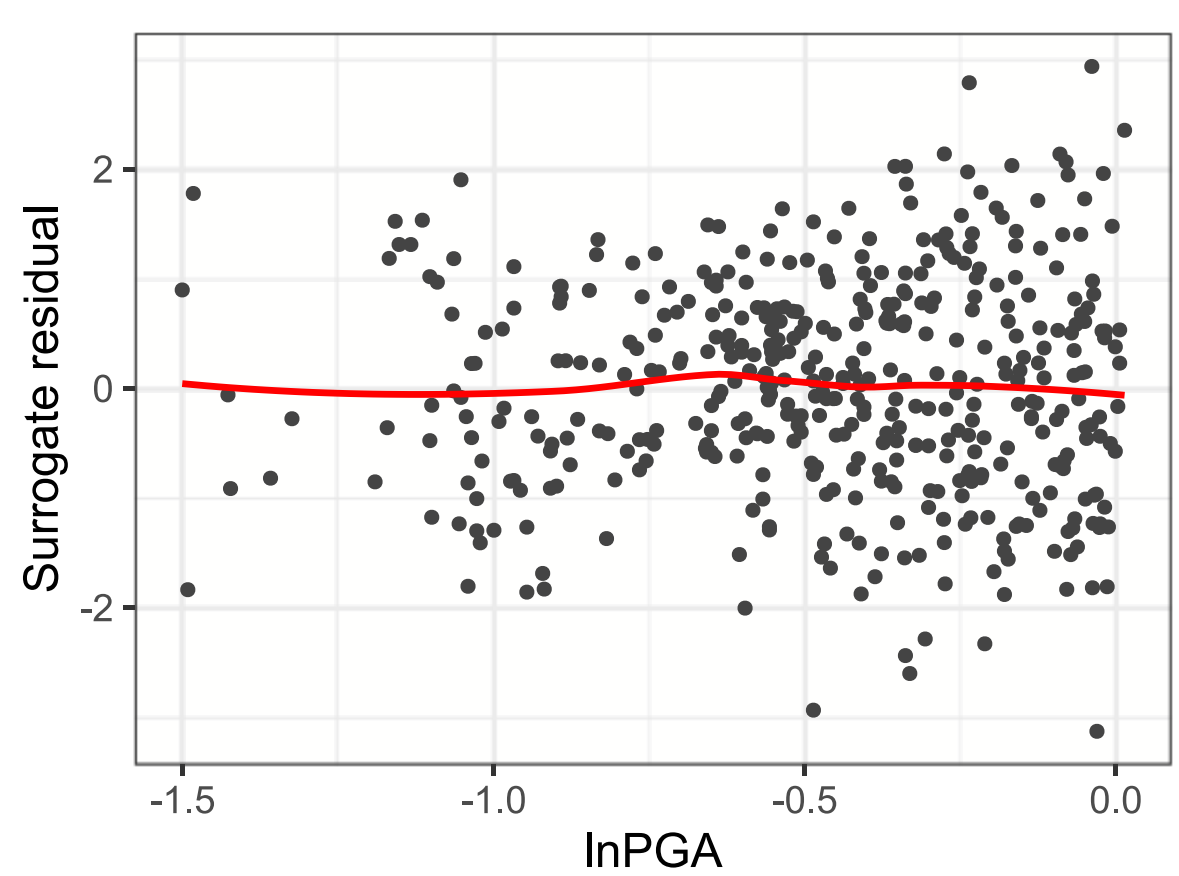

(a) Residual-vs-covariate plot

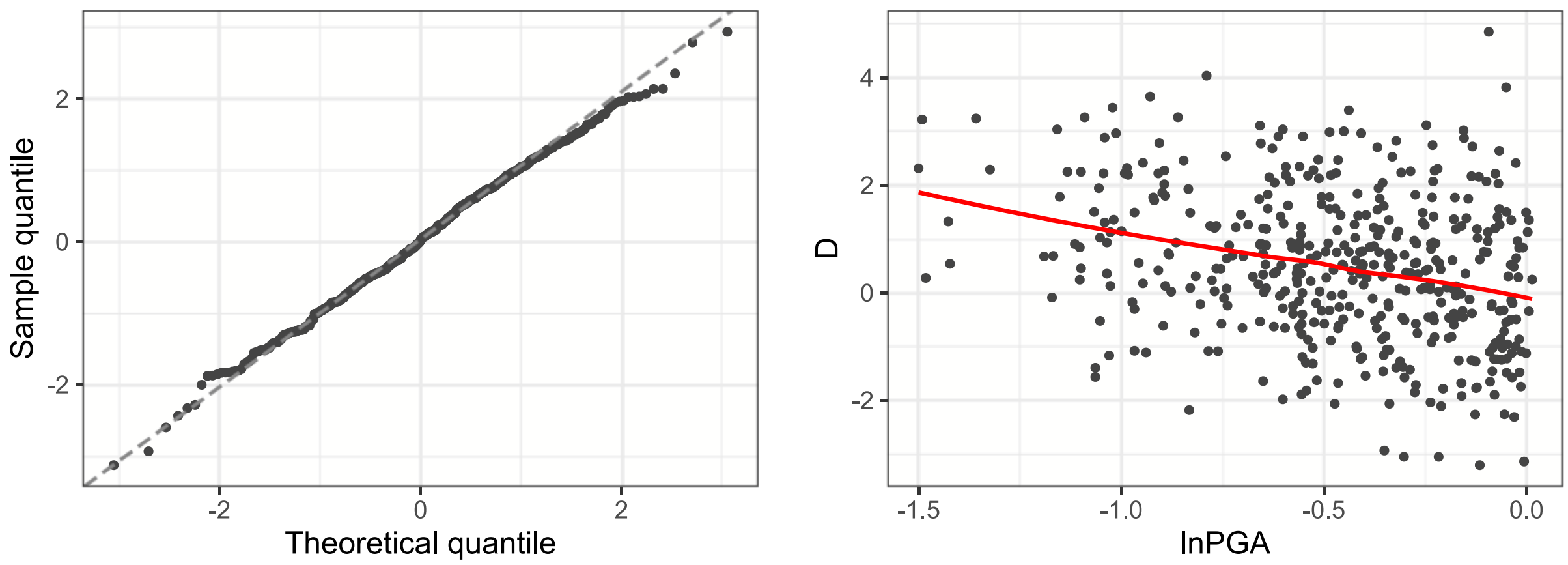

(b) Q-Q plot of the residuals (c) Scatterplot of $D = S_2 - S_1$ vs. covariate

Figure 2 Residuals and diagnostics plot for Cumulative probit regression

(2) Comparison and selection for different modeling strategies

Drawing upon the three types of models and their two extensions outlined in Section 2, this study constructs a series of ordinal regression models for the Wenchuan earthquake bridge damage dataset. The analysis follows specific schemes: (1) constructing three basic regression models based on the fundamental concepts of the model types, adhering to the assumptions of homoscedasticity and proportionality; (2) developing three heteroscedastic models from the basic ones, incorporating variance heterogeneity; (3) relaxing the parallel regression assumption in the Sequential and Adjacent Category models, considering heteroscedasticity and category-specific effects; (4) modifying the Sequential and Adjacent Category models to discard heteroscedasticity while retaining category-specific effects; and (5) for reference, treating damage state as nominal categorical data by employing multinomial logistic regression[56,57]. These eleven models, designated "fit_sc01" to "fit_sc11", are detailed in Table 1. All of the models assume first-order linear functions for mean structures with respect to independent variable (lnPGA), and using a probit link function except in the multinomial logistic model.

The models were fitted using the "Bayesian Regression Models using Stan" (brms) R package [58]. For each model, a weakly informative prior, specifically $\mathcal{N}(0,10)$, was selected for $\beta$, $\gamma$ and $\tau_k$. Four chains were executed, each comprising 1000 warm-up iterations and an additional 1000 iterations. Algorithm convergence was verified through visual inspections, such as traceplots, and by evaluating the Rhat statistic. Due to space constraints, this paper omits detailed parameter estimation results for the all of the models. Specifically, it compares the first model's results with those obtained using the frequentist approach (Cumulative probit regression), as detailed in the previous section, see Table 2. The results show that the Frequentist estimates are pretty close to the Bayesian estimates with naive priors. The consistency observed between these results validates the reasonableness and feasibility of employing the Bayesian method.

Table. 1 Comparison of various ordinal regression models

| Model name | Model description | Number of parameters | elpd_loo | elpd_diff | se_diff | Model ranking |
|---|---|---|---|---|---|---|
| fit_sc01 | Cum | 5 | -647.7 | -9.6 | 4.9 | 9 |
| fit_sc02 | Seq | 5 | -648.4 | -10.2 | 5.1 | 10 |
| fit_sc03 | Acat | 5 | -649.3 | -11.1 | 5.1 | 11 |
| fit_sc04 | Cum + VH | 6 | -646.3 | -8.1 | 4.8 | 7 |
| fit_sc05 | Seq + VH | 6 | -646.9 | -8.8 | 5.1 | 8 |
| fit_sc06 | Acat + VH | 6 | -644.4 | -6.3 | 4.6 | 6 |
| fit_sc07 | Seq + VH + CS | 9 | -638.4 | -0.2 | 0.8 | 2 |
| fit_sc08 | Acat + VH + CS | 9 | -639.7 | -1.5 | 1.9 | 3 |
| fit_sc09 | Seq + CS | 8 | -638.2 | 0 | 0 | 1 |
| fit_sc10 | Acat + CS | 8 | -639.9 | -1.7 | 1.5 | 4 |
| fit_sc11 | Mlogit | 8 | -640 | -1.8 | 1.5 | 5 |

Note: Cum= Cumulative model, Seq=Sequential model, Acat= Adjacent category model, VH= variance heterogeneity, CS= category-specific effects, Mlogit= multinomial logistic model

Table. 2 Comparison of parameter estimations in Cumulative probit models: Frequentist vs. Bayesian Approaches

| term | MASS::polr | | | | brms::brm | | | |
|---|---|---|---|---|---|---|---|---|
| | estimate | std.error | z_value | pr_z | estimate | std.error | conf.low | conf.high |
| $\tau_1$ | -1.617 | 0.111 | -14.546 | NA | -1.620 | 0.109 | -1.836 | -1.412 |
| $\tau_2$ | -1.000 | 0.101 | -9.909 | NA | -1.001 | 0.101 | -1.197 | -0.805 |
| $\tau_3$ | -0.082 | 0.096 | -0.856 | NA | -0.083 | 0.097 | -0.277 | 0.098 |
| $\tau_4$ | 0.623 | 0.102 | 6.083 | NA | 0.627 | 0.101 | 0.430 | 0.825 |
| $\beta$ | 1.549 | 0.168 | 9.236 | <0.001 | 1.554 | 0.170 | 1.227 | 1.897 |

As outlined in Section 2.4, this study employs the leave-one-out cross-validation (LOOCV) method, using the 'loo' R package, to evaluate and compare the eleven models. This approach estimates elpd_loo, the Bayesian LOO estimate of expected log pointwise predictive density for each model. While the magnitude of elpd_loo for an individual model is not intrinsically meaningful for evaluation and selection, the critical aspect is the comparison of elpd_loo values across different models. This involves calculating elpd_diff, the difference in expected predictive accuracy represented by elpd_loo variations, and se_diff, the standard error of this difference, which addresses uncertainty in elpd estimates. Vehtari (2023) suggest a rule of thumb for model comparison [59]: an elpd_diff less than 4 is considered a minor difference. Conversely, an elpd_diff exceeding 4 should be assessed against its se_diff. Generally, an elpd_diff more than double the se_diff indicates statistical significance. When comparing multiple models, elpd_diff and se_diff values result from pairwise comparisons with the model exhibiting the highest elpd. Consequently, the highest elpd model has an elpd_diff of 0 (indicating no difference from itself) and other models show negative elpd_diff values.

Table 1 presents the elpd_loo estimates and model rankings for the 11 models, along with the corresponding elpd_diff and se_diff values for each model relative to the one with the highest elpd. Analysis of the elpd_loo results reveals several insights: initially, the three basic models (fit_sc01~fit_sc03) exhibit similar performance, with the Cumulative model outperforming others. When considering variance heterogeneity, the performance of the three heteroskedastic models (fit_sc04~fit_sc06) shows improvement over the basic models to varying degrees. Notably, the Adjacent Category model demonstrates the most significant enhancement. Incorporating category-specific effects into the Sequential and Adjacent Category models, on top of heteroscedasticity considerations (fit_sc07~fit_sc08), leads to a substantial improvement in model performance. Further, discarding the heteroscedasticity assumption while maintaining category-specific effects (fit_sc09~fit_sc10) results in slight performance fluctuations. Lastly, the multinomial logistic model (fit_sc11) slightly underperforms compared to the preceding four models. These findings align with the previously stated diagnostic conclusions, indicating that for this dataset, the heteroscedasticity effect is relatively minor, while the category-specific effect is pronounced.

Among the eleven models evaluated, the Sequential model that incorporates the category-specific effects (fit_sc09) emerges as the best performer, indicated by the highest elpd_loo value, and thus is established as the baseline with an elpd_diff of 0. According to the aforementioned rule of thumb, the latter five models (fit_sc07~fit_sc11) exhibit no significant differences among themselves, each having an absolute elpd_diff value below 4. In contrast, the first six models (fit_sc01~fit_sc06) show relatively poor performance, with their absolute elpd_diff values exceeding 4. A detailed analysis of se_diff reveals that for the initial three basic models, the absolute elpd_diff is generally greater than or approximately twice the se_diff, leading to a definitive conclusion: the Sequential model incorporating the category-specific effects is markedly superior to the first three basic models.

(3) Seismic fragility results

This study presents the seismic fragility curves derived from both the commonly used basic Cumulative model (fit_sc01) and the Sequential model incorporating category-specific effects (fit_sc09), as illustrated in Figures 3 and 4. Here, (a) represents the exceeding probabilities of the damage states, and (b) the occurrence probabilities of the damage states. Notably, to align with the theoretical framework in Section 2 and for ease of plotting and comparison, this paper focuses on exceeding probabilities, i.e. $\mathrm{P}(DS > k \mid IM = x)$, diverging from the use of reaching or exceeding probabilities $\mathrm{P}(DS \geq k \mid IM = x)$ in traditional seismic fragility analysis. In the context of fragility, the two definitions diverge by a single tier: specifically, the probability of exceedance for complete damage is set at 0 in the first definition, while in the second, the probability of reaching or exceeding none damage is fixed at 1.

A comparative examination of Figures 3 and 4 demonstrates significant disparities in the probabilistic predictions of the two models. Employing Bayesian statistical inference, as opposed to the traditional frequentist approach, facilitates the intuitive and convenient provision of interval estimates alongside point estimates. This method highlights the advantages of Bayesian techniques, especially in integrating epistemic uncertainty into model forecasts. Notably, the Sequential model incorporating category-specific effects, with its eight parameters, displays marginally wider credible intervals, a consequence of the heightened uncertainty associated with parameter estimation. This observation aligns with the concept that an increased number of parameters within a model tends to expand the posterior credible intervals, mirroring the uncertainty embedded in the parameter estimations.

To further elucidate the differences in damage probability predictions, facet plots were created by dissecting subsection (b) in Figures 3 and 4. These plots focus on specific damage probability outcomes across six levels of ground motion intensity, visually represented through bar charts, as Figures 5. The findings indicate that for the two extreme scenarios—none damage and complete damage—the predictive trends of both models align closely. Specifically, the probability of none damage decreases monotonically with rising ground motion intensities, and the outcomes from both models are nearly identical. On the other hand, the probability of complete damage escalates monotonically with increasing ground motion intensities, with the Cumulative model displaying a propensity to underestimate the probability of complete damage at elevated ground motion levels. For the intermediate three damage states, overlapping results were observed. For instance, under PGA values of 0.2g or 0.4g, the Cumulative model tends to overestimate the probability of minor damage while underestimating moderate damage. Conversely, at higher intensities (PGA=1.0g or 1.2g), it overestimates moderate and major damage probabilities. It is important to recognize that these findings are specific to the dataset used and do not represent universal trends. Model preferences and rankings may vary with different datasets, potentially amplifying or reducing the observed differences.

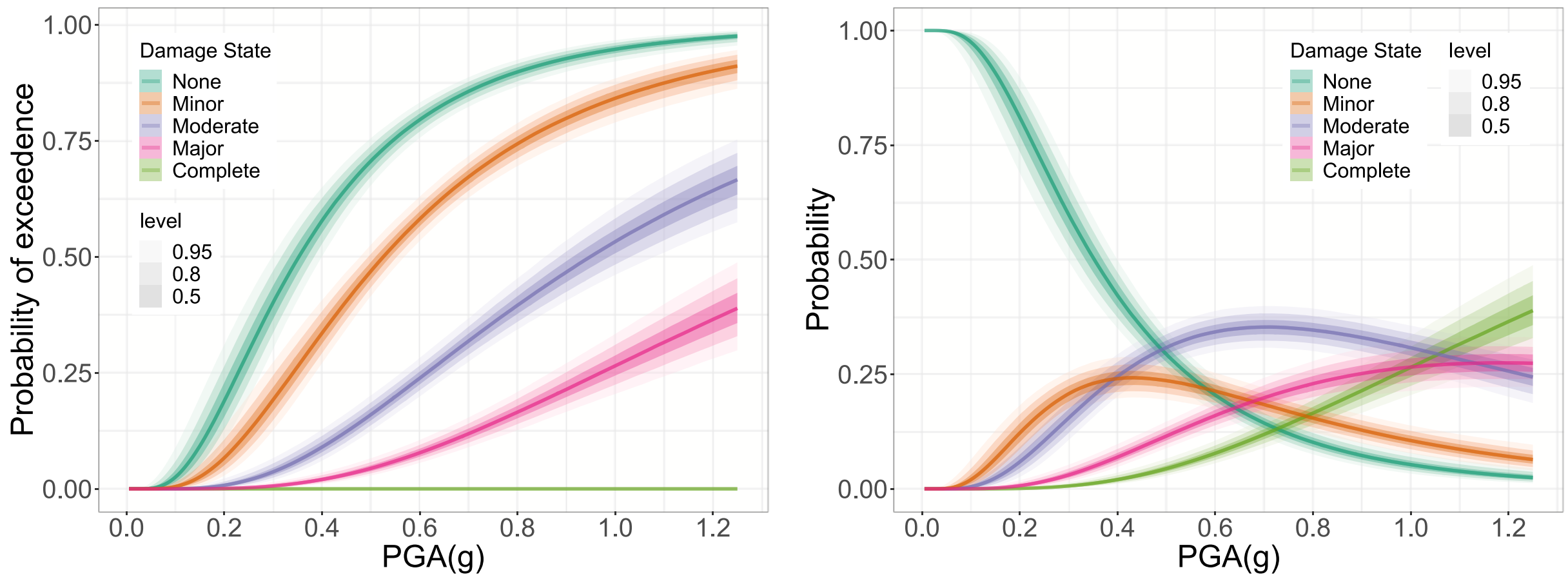

(a) Exceeding probabilities of the damage states (b) Occurrence probabilities of the damage states

Figure 3 Fragility results of Cumulative model

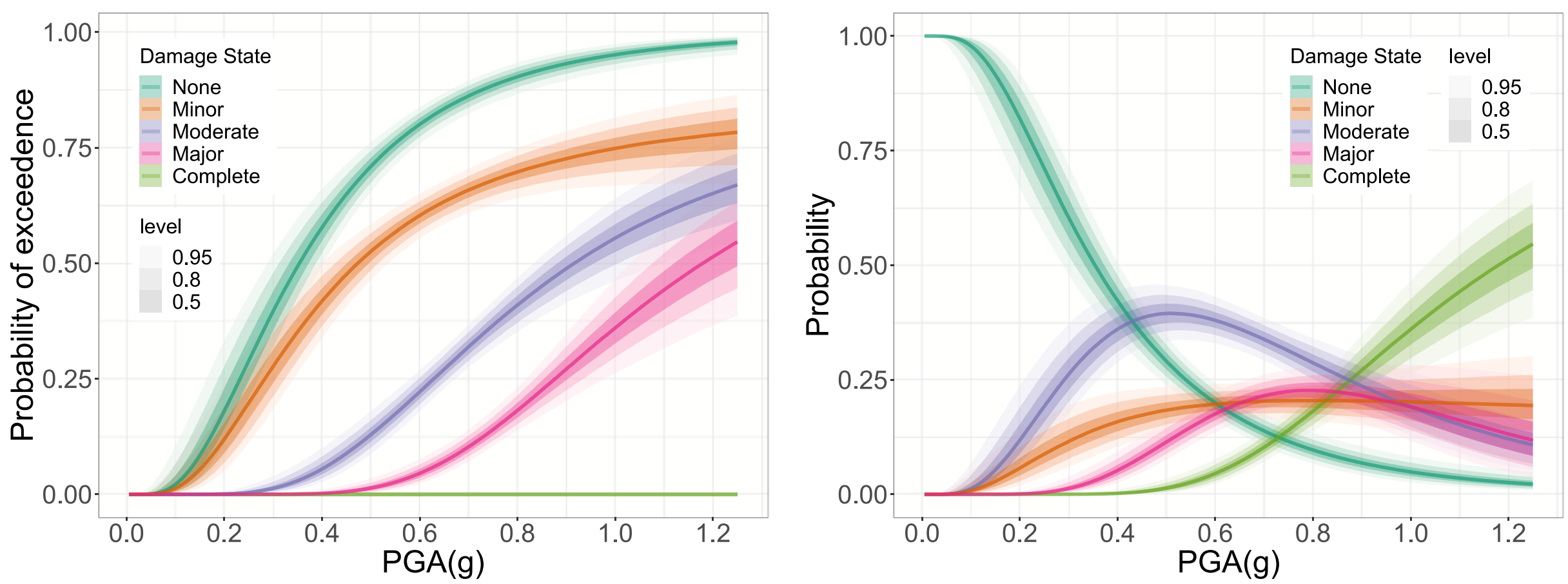

(a) Exceeding probabilities of the damage states (b) Occurrence probabilities of the damage states

Figure 4 Fragility results of Sequential model incorporating category-specific effects

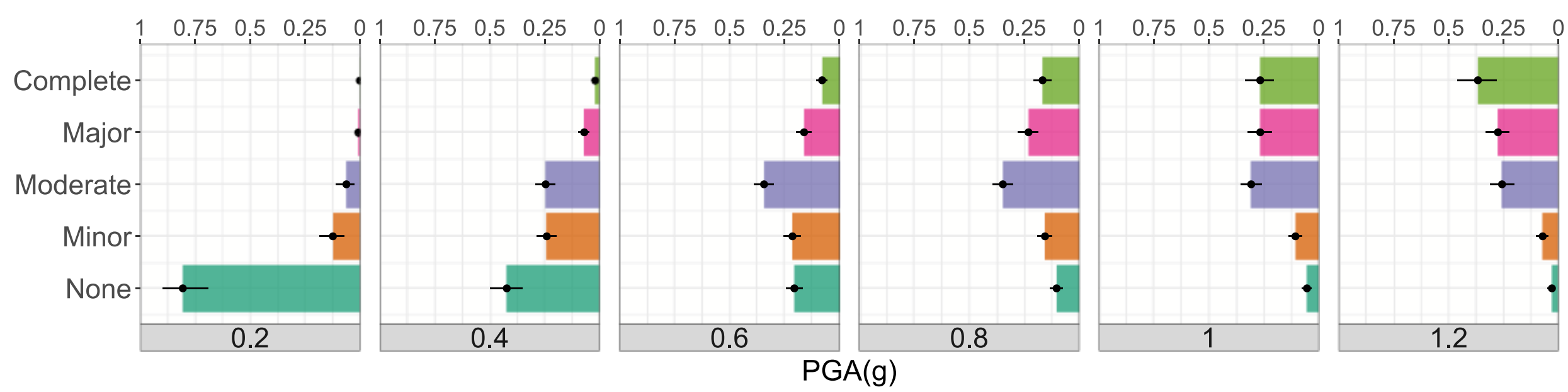

(a) Cumulative model

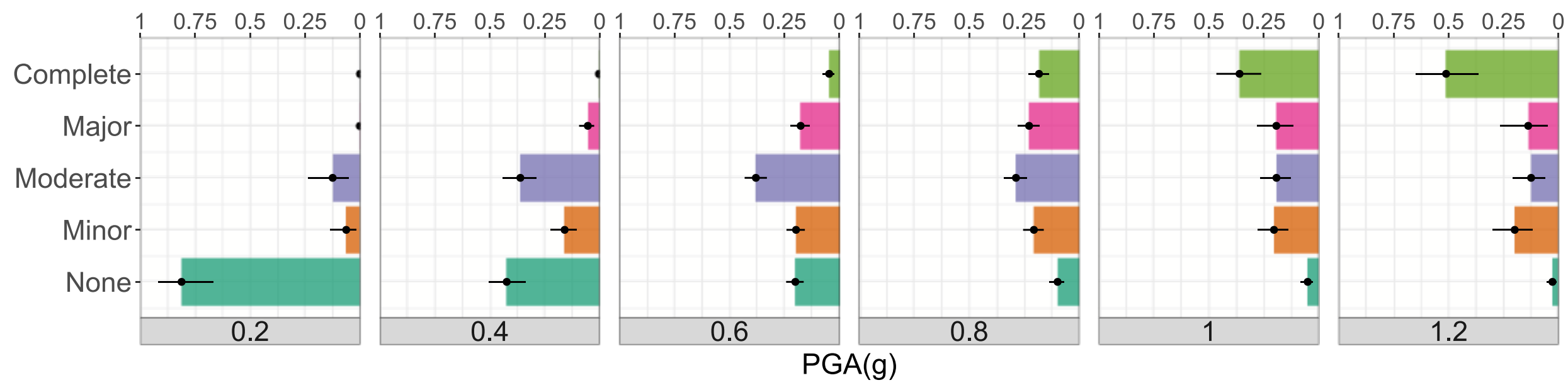

(b) Sequential model incorporating category-specific effects

Figure 5 Facet plots of the damage probabilities

## 4. DISCUSSION AND PERSPECTIVES

### *4.1 Historical notes*

Historically, Shinozuka was the pioneer in introducing the use of the Sequential model for developing fragility curves that encompass multiple damage states [15]. In his seminal work, he critically questions the universal applicability of the lognormal distribution function in accurately characterizing the exceedance probability for all damage states. He also provides a comprehensive explanation of the sequential modeling process in the context of seismic fragility construction. Notably, while Shinozuka does not present an extensive mathematical derivation, he ultimately opts for lognormal distribution function to model fragility across various damage states, citing mathematical expediency and computational ease as his reasons. However, his presentation and discussion offer a novel perspective for the construction of seismic fragility models. For clarity and ease of subsequent derivation, a relevant excerpt from the report is quoted below:

*"Although the assumption of lognormal distribution functions with identical log-standard deviation satisfies the requirement just mentioned, this is not sufficient to theoretically justify the use of lognormal distribution functions for fragility curves associated with all states of damage. In this regard, it is possible to develop a conditional fragility curve associated with each state of damage. This is achieved by implementing the following three steps: first, consider the (unconditional) fragility curve for a state of "at least minor" damage. Second, develop the conditional fragility curve for bridges with a state of damage one rank severer, i.e., "at least moderate" damage. This conditional fragility curve is constructed for the bridges in a state of "at least moderate" damage, considering only those bridges in the "at least minor" state of damage. Finally, the conditional fragility value for the "at least moderate" state of damage is multiplied by the unconditional fragility value for the "at least minor" state of damage at each value of ground motion intensity to obtain the unconditional fragility curve for the "at least moderate" state of damage. Sequentially applied, this three-step process will produce a family of four fragility curves for "at least minor", "at least moderate", "at least major" and "collapse" (in the case of Caltrans' bridges considered in this study) which will not intersect. The fragility curve for "at least minor" state of damage is unconditional to begin with since the state of damage one rank less severe is the state of "at least no" damage which is satisfied by each and every bridge of the entire sample of bridges."*

Expressing the aforementioned passage in statistical terms yields another kind of elaboration on the Sequential model. Rather than modeling the probability that a sequential process halts at category $k$, i.e. Equation (7), it is also possible to model the probability of the sequential process to continue beyond category $k$:

$$\mathrm{P}(Y > k|Y \geq k, \boldsymbol{x}) = 1 - \mathrm{P}(Y = k|Y \geq k, \boldsymbol{x}) = 1 - F(\tau_k - \boldsymbol{x}^T\beta), \qquad k = 1,2 \dots, K-1 \tag{19}$$

Based on the previous assumptions and derivations, we can determine the conditional exceedance probability of a damage state at a given ground motion intensity $IM = x$, as shown in Equation (20).

$$\mathrm{P}(DS > k|DS \geq k, IM = x) = 1 - \Phi(\tau_k - \beta\ln(x)) = \Phi(\beta\ln(x) - \tau_k) \tag{20}$$

At this stage of seismic fragility analysis, we can bypass the need to calculate exceedance probabilities using equations (7) and (8) by accumulating unconditional damage probabilities. Instead, a more efficient approach, as described by Shinozuka, is to directly calculate the results using equation (20) through concatenated multiplication.

$$\mathrm{Fr}_k(IM = x) = \mathrm{P}(DS > k|IM = x) = \prod_{j=1}^{k} \mathrm{P}(DS > j|DS \geq j, IM = x) = \prod_{j=1}^{k} \Phi\big(\beta\ln(x) - \tau_j\big) \tag{21}$$

The aforementioned derivation elucidates why Sequential model is often termed "Continuation ratio" models, as it emphasize the process of progression from one stage to the subsequent one [19,25,28]. Similarly, the model derived within the Section 2 is referred to as "Stopping ratio" models, where the conditional probability is the probability of stopping in a particular stage or outcome category [60]. Both are equivalent to each other for symmetric link functions such as probit or logit but will differ for asymmetric ones such as cloglog.

*4.2 Implications for analytical seismic fragility modeling*

Over the past three decades, the proliferation of computer technology and numerical analysis methods has led to the extensive application of analytical seismic fragility models in diverse engineering disciplines. Unlike empirical models that depend on realistic seismic data, analytical models enable a more focused assessment of specific structures' performance. The precision of structural numerical analysis and the acquisition of large-scale sample data facilitate the development of more accurate and broadly applicable seismic fragility models. Analytical fragility modeling typically employs one of two strategies: the first decouples seismic fragility into a convolution of two intermediate terms within the performance-based earthquake engineering framework, representing the probability of seismic demand at a given intensity and the probability of exceeding a certain damage state given the seismic demand, respectively. These are known as the probabilistic seismic demand model and the capacity fragility model, whose combination yields a fragility curve. Alternatively, if the process of constructing a seismic demand model is omitted, researchers may directly compile a dataset consisting of ground motion intensity measures and damage state samples to establish a fragility model, as outlined in this study.

In the first approach, the methodology entails conducting batch numerical simulations, predominantly nonlinear dynamic time-history analyses, tailored to the specific subject of study. Subsequent to generating seismic demand samples for the structures and integrating them with intensity measure samples, regression analysis is employed to develop the corresponding probabilistic seismic demand model. Commonly, the academics adopts Cornell's first-order linear model in logarithmic space [61], premised on the assumption that seismic demand, in response to specified ground motion intensity, conforms to a lognormal distribution $\ln(D) \sim \mathcal{N}(\ln(S_D), \beta_D^2)$, with its median value demonstrating a power-law relation to the seismic intensity. Logarithmic transformation of the equation facilitates the derivation of a first-order linear model in logarithmic space, as Equation (22), enabling the estimation of coefficients and variance through the least squares method. Integrating this with the capacity model, which typically involves setting limit states presumed to adhere to lognormal distributions $\ln(C) \sim \mathcal{N}(\ln(S_C), \beta_C^2)$, streamlines the process to deduce exceedance probabilities for various damage states, as illustrated by Equation (23).

$$\ln(D) = \ln(\alpha_0) + \alpha_1\ln(IM) + \sigma\varepsilon \tag{22}$$

$$P[D \geq C \mid IM] = \Phi\left(\frac{\ln\left(\frac{S_D}{S_C}\right)}{\sqrt{\beta_D^2 + \beta_C^2}}\right) = \Phi\left(\frac{\ln(IM) - \frac{\ln(S_C) - \ln(\alpha_0)}{\alpha_1}}{\frac{\sqrt{\beta_{D|IM}^2 + \beta_C^2}}{\alpha_1}}\right) = \Phi\left(\frac{\ln(IM) - \ln(\theta_{Fr})}{\beta_{Fr}}\right) \tag{23}$$

The analysis presented indicates that a fragility function can be reduced to a closed form through a conventional two-parameter lognormal distribution function, provided that both seismic demand and capacity adhere to a lognormal distribution and that seismic demand exhibits a first-order linear relationship with seismic intensity in logarithmic space. While Cornell highlighted the significance of this logarithmic linear relationship, positing that seismic intensity measures should fall within a specific range, this consideration has often been overlooked in the practical application of this model to seismic fragility analysis. In fact, the heteroscedasticity of the logarithmic demand has been noted by some scholars in the existing literature [38,39]. A potential remedy, akin to the approach in this study for addressing variance heterogeneity in fragility modeling, involves incorporating the variance component into the demand model's regression analysis. In other words, simultaneous estimation is achieved by considering both the mean and variance of the logarithm of seismic demand as functions of the intensity measure, e.g. $E(\ln D) = f(IM)$, $Var(\ln D) = \exp\big(g(IM)\big)$. This method also allows for the exploration of both mean and variance's nonlinear trends, either parametrically or non-parametrically, enhancing the depiction of seismic demand's probabilistic attributes across various seismic intensities. By accurately determining the mean and variance and reintegrating them into the model, a refined fragility function emerges, however it will be diverging from the traditional paradigm of lognormal distribution function.

It is important to emphasize that, for the construction of multi-damage state fragility models, special attention should be paid to the setting of the capacity models in the equation. The definition of limit states represents the mapping connection between the magnitude of demand measures and actual seismic damage sustained by structural components. In most studies, the setting of these values depends on the empirical assumptions of researchers. While the fragility equation remains valid if the uncertainty of multiple limit states is overlooked or if it is assumed that all limit states share an identical logarithmic standard deviation, issues arise when considering the variability in uncertainty across different limit states. For instance, assigning varied logarithmic standard deviations to different limit states can lead to intersecting fragility curves upon reintegration into the equation. This outcome, akin to the Cumulative model addressing category-specific effects as discussed in Section 2, defies logical consistency.

In such instances, a viable solution involves employing the secondary approach outlined previously, which entails acquiring seismic demand samples across various seismic intensities via nonlinear time-history analysis, alongside limit state samples generated on predefined capacity models. In the capacity sampling phase, it is also permissible to relax the assumption that limit states adhere strictly to lognormal distributions, and the differences in uncertainty of limit states can be taken into account during the sampling process. Noteworthy, it's crucial to maintain a sequential arrangement of the limit state samples for each sampling, potentially through the application of an appropriate correlation coefficient among different limit states or the acceptance-rejection method. Upon gathering the limit state samples and juxtaposing them with the seismic demand samples, damage state samples emerge, facilitating the creation of a matched set of intensity measures and damage states. Consequently, the fragility models can be developed utilizing the methodologies delineated in this study.

### *4.3 Model selection in practice and further extensions*

Within the context of the modeling framework presented in this paper, it is feasible to incorporate additional ground motion intensity measures as independent variables to enhance the precision of fragility forecasts. Two considerations are necessary at this point: First, the choice of intensity measures is critical. While introducing a broader range of indicators enriches the dataset with more detailed ground motion insights, it often leads to significant correlations among them, posing a challenge due to multicollinearity in regression analysis. This issue can potentially destabilize the estimation of model parameters. Techniques such as Ridge or Lasso regression offer solutions by mitigating multicollinearity and facilitating variable selection. Second, it is essential to ensure that seismic fragility models align with the framework of Performance-Based Earthquake Engineering in the specific application. Traditional probabilistic seismic hazard models typically do not support joint distribution forecasts for multiple seismic intensity measures, which can restrict the integration of multi-predictor fragility models in seismic risk or resilience assessment. In such cases, reverting to the conventional single-predictor framework and selecting a more suitable ground motion intensity measure, combined with an improved modeling approach, might be a preferable alternative.

In specific contexts, the link function and the mean structure's nonlinearity can influence seismic fragility outcomes. Given that the dataset chosen for this study closely conforms to the initial assumptions, an extensive discussion on this topic was omitted. However, when faced with these challenges, the methodologies outlined in this paper offer guidance for model diagnostics and comparisons. This includes selecting the most suitable link function and incorporating nonlinearity through approaches like Splines or Gaussian processes, thereby improving the predictive precision of fragility models.

From the perspective of research objectives, models can be categorized into two main types: diagnostic, which are aimed at deciphering underlying principles, and predictive, which forecast system behaviors based on these principles. Hence, the qualification of fragility models is thus contingent upon the researcher's purpose. As mentioned in the introduction of this paper, the adoption of a two-parameter lognormal distribution function as the foundational model for fragility presents several benefits, including simplicity, efficiency, and ease of interpretation and comparison, making it a preferred option in numerous contexts. Nonetheless, it is imperative for researchers to grasp the foundational logic and assumptions of the model. Should the collected data diverge from these assumptions (such as homoscedasticity or the parallel regression assumption), and the research focus is on the precise prediction of damage probabilities, then the constructed model and its details warrant reevaluation. One of the initial intentions of this paper is to improve the precision of probabilistic predictions by evaluating and selecting among various modeling approaches under identical conditions (i.e., utilizing the same predictor and dataset). Viewed through this lens, the fragility models discussed herein fall into the predictive category.

From the perspective of generation mechanisms, models can generally be categorized into law-driven and data-driven types. The latter becomes a reasonable choice when physical principles are ambiguous and many influencing factors occur randomly. In earthquake engineering, accurately predicting a structure's damage state is often unfeasible due to substantial uncertainties and the limitations of using a single ground motion intensity measure to capture the full extent of seismic motion. This challenge has led to the adoption of the probabilistic Performance-Based Earthquake Engineering framework and the development of the fragility concept. This paper elucidates that seismic fragility modeling essentially constitutes a unique form of regression model aimed at the probabilistic prediction of specific categories under defined conditions, aligning well with the data-driven approach. It is important to highlight that this research is conducted within a conventional statistical framework. However, numerous machine learning techniques, such as support vector machines (SVMs) [62] and Gaussian processes [63], have been developed and employed to address ordinal regression challenges. The adaptation of these methodologies to seismic fragility modeling warrants additional exploration.

## 5. CONCLUSION

This study systematically explores the application of a range of ordinal regression models for fitting seismic fragility curves, providing an alternative to the conventional reliance on the lognormal distribution function. By examining the distinctions and connections among different ordinal regression models, including Cumulative, Sequential, and Adjacent Category models, the research highlights the potential advantages of using models that account for category-specific effects and variance heterogeneity. Both frequentist and Bayesian methods were employed for statistical inference, alongside a surrogate residual method for model evaluation. Model comparison and assessment were conducted using an enhanced leave-one-out cross-validation technique, offering a detailed diagnostic framework for model selection.

A case study using bridge damage data from the Wenchuan earthquake demonstrates the practical application of these methodologies. Within this dataset, the Sequential model incorporating category-specific effects showed improved predictive precision compared to traditional Cumulative models. This empirical application offers insights into the process of evaluating and selecting seismic fragility models, emphasizing the importance of addressing model-specific assumptions and limitations to improve the accuracy and reliability of seismic fragility assessments.

While this study focuses on bridge damage fragility assessments, its methodological approach has potential implications for broader seismic engineering and disaster risk management practices. Future research directions could include:

1. Extending the refined fragility models to seismic risk assessment and resilience evaluations within the framework of Performance-Based Earthquake Engineering. This could involve analyzing how different fragility models affect risk and resilience metrics, with particular attention to uncertainty propagation.
2. Investigating the evaluation and selection of ground motion intensity measures based on seismic fragility models, potentially improving the accuracy and applicability of fragility assessments across various structural types and seismic scenarios.

By addressing these areas, future studies can build on this research to further develop seismic fragility analysis, contributing to more refined risk assessments and resilience planning in the context of seismic performance and infrastructure protection.

## APPENDIX

To facilitate the derivation, we rename the two variables in Equations (9) and (10) as follows:

$$F_k := F(\tau_k - \boldsymbol{x}^T\beta), \qquad \text{for } k = 1,2 \dots, K-1$$

$$P_k := \mathrm{P}(Y = k|\boldsymbol{x}), \qquad \text{for } k = 1,2 \dots, K$$

The original problem is transformed into a scenario where, given $K$ probabilities $P_1, P_2, \dots, P_K$ and $K-1$ ratios $F_1, F_2, \dots, F_{K-1}$, with the constraint $\sum_{k=1}^{K} P_k = 1$, and we have the following ratio relationships:

$$\frac{P_1}{P_1+P_2}=F_1$$
$$\frac{P_2}{P_2+P_3}=F_2$$
$$\vdots$$
$$\frac{P_{K-1}}{P_{K-1}+P_K}=F_{K-1}$$

The object is to derive the recursive expression for $P_k$ in terms of $F_k$.

The detailed derivation of Equation (10) is as follows:

（1）From the definition of the ratio relationships, we obtain:

$$\begin{gathered} P_1 = F_1(P_1+P_2) \\ P_2 = F_2(P_2+P_3) \\ \vdots \\ P_{K-1} = F_{K-1}(P_{K-1}+P_K) \end{gathered} \tag{a1}$$

Starting from $P_1$, we can recursively express:

$$\begin{gathered} P_2 = \frac{P_1(1-F_1)}{F_1} \\ P_3 = \frac{P_2(1-F_2)}{F_2} \\ \vdots \\ P_K = \frac{P_{K-1}(1-F_{K-1})}{F_{K-1}} \end{gathered} \tag{a2}$$

Given these relationships, we can write the recursive formula:

$$P_{k+1} = \frac{P_k(1-F_k)}{F_k}, \qquad \text{for } k = 1,2,\dots,K-1 \tag{a3}$$

This means $P_{k+1}$ is the product of $P_k$ and $(1-F_k)/F_k$.

（2）Furthermore, by recursively applying this form, we can express $P_k$ as a function of $P_1$. Starting from $P_1$:

$$\begin{gathered} P_2 = \frac{P_1(1-F_1)}{F_1} \\ P_3 = \frac{P_2(1-F_2)}{F_2} = \frac{P_1(1-F_1)(1-F_2)}{F_1F_2} \\ \vdots \\ P_K = \frac{P_1\prod_{j=1}^{K-1}(1-F_j)}{\prod_{j=1}^{K-1}F_j} \end{gathered} \tag{a4}$$

Hence,

$$P_k = \frac{P_1\prod_{j=1}^{k-1}(1-F_j)}{\prod_{j=1}^{k-1}F_j}, \qquad \text{for } k = 2,\dots,K \tag{a5}$$

（3）Since $P_1$ is not directly known, we need to solve for $P_1$ using $\sum_{k=1}^{K}P_k = 1$.

$$\begin{gathered} P_1 + \frac{P_1(1-F_1)}{F_1} + \frac{P_1(1-F_1)(1-F_2)}{F_1F_2} + \cdots + \frac{P_1\prod_{j=1}^{K-1}(1-F_j)}{\prod_{j=1}^{K-1}F_j} = 1 \\ P_1\left(1 + \frac{1-F_1}{F_1} + \frac{(1-F_1)(1-F_2)}{F_1F_2} + \cdots + \frac{\prod_{j=1}^{K-1}(1-F_j)}{\prod_{j=1}^{K-1}F_j}\right) = 1 \end{gathered} \tag{a6}$$

To facilitate the normalization of the denominator in subsequent derivations, we introduce $F_K$ and let it always be 1, i.e., $F_K \equiv 1$. We can also add the term $\frac{\prod_{j=1}^{K}(1-F_j)}{\prod_{j=1}^{K}F_j}$ in the parentheses, and since $F_K = 1$, this term is always 0.

We then define:

$$S = 1 + \frac{(1-F_1)}{F_1} + \frac{(1-F_1)(1-F_2)}{F_1F_2} + \cdots + \frac{(1-F_1)(1-F_2)\cdots(1-F_{K-1})}{F_1F_2\cdots F_{K-1}} + \frac{(1-F_1)(1-F_2)\cdots(1-F_K)}{F_1F_2\cdots F_K}$$
$$= \frac{\prod_{j=1}^{K} F_j + (1-F_1)\prod_{j=2}^{K} F_j + (1-F_1)(1-F_2)\prod_{j=3}^{K} F_j + \cdots + \prod_{j=1}^{K}\left(1-F_j\right)}{\prod_{j=1}^{K} F_j}$$

So,

$$P_1 = \frac{1}{S} = \frac{\prod_{j=1}^{K} F_j}{\sum_{r=1}^{K+1}\left(\prod_{j=1}^{r-1}\left(1-F_j\right)\prod_{j=r}^{K} F_j\right)} \tag{a7}$$

Where the denominator sum comprises $K+1$ parts, from $r = 1$ to $r = K+1$

- When $r$ = 1, the preceding product is empty, so it is 1, and the subsequent product is $\prod_{j=1}^{K} F_j$
- When $r$ = 2, the preceding product is $1 - F_1$ , and the subsequent product is $\prod_{j=2}^{K} F_j$
- And so on, until $r = K+1$, where the preceding product is $\prod_{j=1}^{K}\left(1-F_j\right)$, and the subsequent product is empty, so it is 1.

（4）Substituting $P_1$ back into the expression (a5) for $P_k$ :

$$P_k = \frac{P_1\prod_{j=1}^{k-1}\left(1-F_j\right)}{\prod_{j=1}^{k-1} F_j} = \frac{\prod_{j=1}^{K} F_j}{\sum_{r=1}^{K+1}\left(\prod_{j=1}^{r-1}\left(1-F_j\right)\prod_{j=r}^{K} F_j\right)}\frac{\prod_{j=1}^{k-1}\left(1-F_j\right)}{\prod_{j=1}^{k-1} F_j}$$
$$= \frac{\prod_{j=1}^{k-1}\left(1-F_j\right)\prod_{j=k}^{K} F_j}{\sum_{r=1}^{K+1}\left(\prod_{j=1}^{r-1}\left(1-F_j\right)\prod_{j=r}^{K} F_j\right)} \tag{a8}$$

Reducing the expression, and we can obtain the formula in the main text (Equation 10):

$$\mathrm{P}(Y = k|\boldsymbol{x}) = \frac{\prod_{j=1}^{k-1}\left(1-F\left(\tau_j-\boldsymbol{x}^T\beta\right)\right)\prod_{j=k}^{K}F\left(\tau_j-\boldsymbol{x}^T\beta\right)}{\sum_{r=1}^{K+1}\left(\prod_{j=1}^{r-1}\left(1-F\left(\tau_j-\boldsymbol{x}^T\beta\right)\right)\prod_{j=r}^{K}F\left(\tau_j-\boldsymbol{x}^T\beta\right)\right)}, \quad \text{for } k = 1,2\ldots,K \tag{10}$$

## DATA AVAILABILITY STATEMENTS

All data and code that support the findings of this study are available from the author upon reasonable request.

## ACKNOWLEDGEMENT

The author thanks the National Natural Science Foundation of China (51308125) and the Natural Science Foundation of Fujian Province (2020J01478) for their financial support.